\documentclass[aps,prl,preprint,tightenlines,superscriptaddress,showpacs,byrevtex]{revtex4}
\usepackage{graphicx} 
\usepackage{dcolumn}  
\usepackage{pstricks}

\graphicspath{{ps}}

\begin{document}


\preprint{\vbox{ \hbox{   }
                 \hbox{BELLE-CONF-0648}
}}

\title{ \quad\\[0.5cm] Measurements of time-dependent $CP$ violation in $B^0 \to \omega K_S^0$, $f_0(980) K_S^0$, $K_S^0 \pi^0$ 
and $K^+ K^- K_S^0$ decays}


\affiliation{Budker Institute of Nuclear Physics, Novosibirsk}
\affiliation{Chiba University, Chiba}
\affiliation{Chonnam National University, Kwangju}
\affiliation{University of Cincinnati, Cincinnati, Ohio 45221}
\affiliation{University of Frankfurt, Frankfurt}
\affiliation{The Graduate University for Advanced Studies, Hayama} 
\affiliation{Gyeongsang National University, Chinju}
\affiliation{University of Hawaii, Honolulu, Hawaii 96822}
\affiliation{High Energy Accelerator Research Organization (KEK), Tsukuba}
\affiliation{Hiroshima Institute of Technology, Hiroshima}
\affiliation{University of Illinois at Urbana-Champaign, Urbana, Illinois 61801}
\affiliation{Institute of High Energy Physics, Chinese Academy of Sciences, Beijing}
\affiliation{Institute of High Energy Physics, Vienna}
\affiliation{Institute of High Energy Physics, Protvino}
\affiliation{Institute for Theoretical and Experimental Physics, Moscow}
\affiliation{J. Stefan Institute, Ljubljana}
\affiliation{Kanagawa University, Yokohama}
\affiliation{Korea University, Seoul}
\affiliation{Kyoto University, Kyoto}
\affiliation{Kyungpook National University, Taegu}
\affiliation{Swiss Federal Institute of Technology of Lausanne, EPFL, Lausanne}
\affiliation{University of Ljubljana, Ljubljana}
\affiliation{University of Maribor, Maribor}
\affiliation{University of Melbourne, Victoria}
\affiliation{Nagoya University, Nagoya}
\affiliation{Nara Women's University, Nara}
\affiliation{National Central University, Chung-li}
\affiliation{National United University, Miao Li}
\affiliation{Department of Physics, National Taiwan University, Taipei}
\affiliation{H. Niewodniczanski Institute of Nuclear Physics, Krakow}
\affiliation{Nippon Dental University, Niigata}
\affiliation{Niigata University, Niigata}
\affiliation{University of Nova Gorica, Nova Gorica}
\affiliation{Osaka City University, Osaka}
\affiliation{Osaka University, Osaka}
\affiliation{Panjab University, Chandigarh}
\affiliation{Peking University, Beijing}
\affiliation{University of Pittsburgh, Pittsburgh, Pennsylvania 15260}
\affiliation{Princeton University, Princeton, New Jersey 08544}
\affiliation{RIKEN BNL Research Center, Upton, New York 11973}
\affiliation{Saga University, Saga}
\affiliation{University of Science and Technology of China, Hefei}
\affiliation{Seoul National University, Seoul}
\affiliation{Shinshu University, Nagano}
\affiliation{Sungkyunkwan University, Suwon}
\affiliation{University of Sydney, Sydney NSW}
\affiliation{Tata Institute of Fundamental Research, Bombay}
\affiliation{Toho University, Funabashi}
\affiliation{Tohoku Gakuin University, Tagajo}
\affiliation{Tohoku University, Sendai}
\affiliation{Department of Physics, University of Tokyo, Tokyo}
\affiliation{Tokyo Institute of Technology, Tokyo}
\affiliation{Tokyo Metropolitan University, Tokyo}
\affiliation{Tokyo University of Agriculture and Technology, Tokyo}
\affiliation{Toyama National College of Maritime Technology, Toyama}
\affiliation{University of Tsukuba, Tsukuba}
\affiliation{Virginia Polytechnic Institute and State University, Blacksburg, Virginia 24061}
\affiliation{Yonsei University, Seoul}
  \author{K.~Abe}\affiliation{High Energy Accelerator Research Organization (KEK), Tsukuba} 
  \author{K.~Abe}\affiliation{Tohoku Gakuin University, Tagajo} 
  \author{I.~Adachi}\affiliation{High Energy Accelerator Research Organization (KEK), Tsukuba} 
  \author{H.~Aihara}\affiliation{Department of Physics, University of Tokyo, Tokyo} 
  \author{D.~Anipko}\affiliation{Budker Institute of Nuclear Physics, Novosibirsk} 
  \author{K.~Aoki}\affiliation{Nagoya University, Nagoya} 
  \author{T.~Arakawa}\affiliation{Niigata University, Niigata} 
  \author{K.~Arinstein}\affiliation{Budker Institute of Nuclear Physics, Novosibirsk} 
  \author{Y.~Asano}\affiliation{University of Tsukuba, Tsukuba} 
  \author{T.~Aso}\affiliation{Toyama National College of Maritime Technology, Toyama} 
% ...
\collaboration{The Belle Collaboration}

\noaffiliation

\begin{abstract}
We present measurements of time-dependent $CP$ asymmetries in 
$B^0 \to \omega K_S^0$, $f_0 (980) K_S^0$, $K_S^0 \pi^0$ and 
$K^+ K^- K_S^0$ based on a sample of 
535 $\times 10^6$ $B\overline{B}$ pairs collected at the $\Upsilon(4S)$
resonance with the Belle detector at the KEKB energy-asymmetric
$e^+ e^-$ collider. 
One neutral $B$ meson is fully reconstructed in one of the specified
decay channels, and the flavor of the accompanying $B$ meson is identified
from its decay products. $CP$-violation parameters for each of the
decay modes are obtained from the asymmetries in the distributions of
the proper-time intervals between the two $B$ decays.
\end{abstract}


\maketitle

\tighten

The Standard Model (SM) decribes $CP$ violation in $B^0$
meson decays using the complex phase of the $3 \times 3$ 
Cabibbo-Kobayashi-Maskawa (CKM) mixing matrix~\cite{ckm}.
$CP$ asymmetries in neutral $B$ meson decays into $CP$ eigenstates 
$f$ exhibit a time-dependent behavior
\begin{equation}
A (\Delta t) = {\cal S}_f \sin (\Delta m_d \Delta t) + 
{\cal A}_f \cos (\Delta m_d \Delta t)
\end{equation}
where ${\cal S}_f$ and ${\cal A}_f$ are the $CP$ violation parameters, 
$\Delta m_d$ the mass difference between the two $B^0$ mass eigenstates, 
$\Delta t$ the difference between the decay time of the signal $B^0$ 
($\overline{B}{}^0$) and of the opposite-side $\overline{B}{}^0$ ($B^0$).
The SM predicts that for most of the decays that proceed
via the quark transitions $b \to s \overline{q} q$ ($q = u, d, s$)
the relations ${\cal S}_f = -\xi_f \sin 2 \phi_1$ and ${\cal A}_f \simeq 0$, 
where $\xi_f = +1 (-1)$ corresponds to $CP$-even (-odd) final states, hold to 
a good approximation~\cite{theory}.
With physics beyond the SM, these decays may receive significant contributions that 
depend on a phase that is different from the SM prediction. A comparison 
of the effective $\sin 2 \phi_1$ values, $\sin 2 \phi_1^{eff}$, with 
$\sin 2 \phi_1$ obtained from the decays governed by the $b \to c \overline{c} s$ 
transition is thus an important test of the SM.
%
%
\par Among the final states studied, $\omega K_S^0$ and $K_S^0 \pi^0$ 
are $CP$-odd modes,
$f_0(980) K_S^0$ is a $CP$-even mode, while
$K^+ K^- K_S^0$ is a mixture of both $\xi_f = -1$ and $+1$. The SM expectation 
for this latter mode is ${\cal S}_f = -(2 f_{+} - 1) \sin 2 \phi_1$, where 
$f_{+}$ is the $CP$-even fraction.
A measurement of $f_+$ was obtained using isospin relation~\cite{cpeven}
with a 357 fb$^{-1}$ data sample and gives 
$f_{+} =  0.93 \pm 0.09 (\rm stat) \pm 0.05 (syst)$. 
%
%
\par Recently, it was found that the direct $CP$ asymmetries in 
$B^0 \to K^+ \pi^-$ and $B^+ \to K^+ \pi^0$ differ significantly~\cite{hfag} 
while they were naively expected to be same~\cite{gronau0}. 
Using the $B^0 \to K_S \pi^0$ result, an additional test to understand the situation 
can be made by comparing the measured ${\cal A}_f$
value and the value predicted by a sum rule~\cite{gronau} using asymmetry
measurements from the other $B \to K\pi$ decays.
%
%
\par Previous measurements of $CP$ asymmetries in $b \to s \overline{q} q$ transitions
have been reported by Belle~\cite{belle_b2s} and 
BaBar~\cite{babar_b2s}. Belle's previously published results of $CP$ 
in $B^0 \to \omega K_S^0$, $f_0 (980) K_S^0$, $K_S^0 \pi^0$
and $K^+ K^- K_S^0$ were based on a 253~fb$^{-1}$ data sample
corresponding to $275 \times 10^6$ $B \overline{B}$ pairs. In this report, we describe
improved measurements incorporating an additional 239~fb$^{-1}$ data sample for
a total of 492~fb$^{-1}$ ($535 \times 10^6$ $B \overline{B}$ pairs).
\par At the KEKB energy-asymmetric $e^+ e^-$ (3.5 on 8.0 GeV) collider, the 
$\Upsilon(4S)$ is produced with a Lorentz boost of $\beta \gamma = 0.425$
nearly along the electron beamline ($z$). Since the $B^0$ and $\overline{B}{}^0$
are approximatively at rest in the $\Upsilon(4S)$ center-of-mass system (cms),
$\Delta t$ can be determined from the displacement in $z$ between the two 
decay vertices: $\Delta t \equiv \Delta z / (\beta \gamma c)$.
%
\par The Belle detector is a large-solid-angle magnetic
spectrometer that consists of a silicon vertex detector (SVD),
a 50-layer central drift chamber (CDC), an array of
aerogel threshold \v{C}erenkov counters (ACC),
a barrel-like arrangement of time-of-flight
scintillation counters (TOF), and an electromagnetic calorimeter (ECL)
comprised of CsI(Tl) crystals located inside
a superconducting solenoid coil that provides a 1.5~T
magnetic field.  An iron flux-return located outside of
the coil is instrumented to detect $K_L^0$ mesons and to identify
muons (KLM).  The detector is described in detail elsewhere~\cite{Belle}.
Two different inner detector configurations were used. For the first sample
of 152 $\times 10^6$ $B\overline{B}$ pairs, a 2.0 cm radius beampipe
and a 3-layer silicon vertex detector were used;
for the latter 383 $\times 10^6$ $B\overline{B}$ pairs,
a 1.5 cm radius beampipe, a 4-layer silicon detector
and a small-cell inner drift chamber were used~\cite{Ushiroda}.
%
\par We reconstruct the following $B^0$ decay modes to measure $CP$
asymmetries: $B^0 \to \omega K_S^0$, $f_0 (980) K_S^0$, $K_S^0 \pi^0$
and $K^+ K^- K_S^0$. We exclude $K^+ K^-$ pairs that are consistent
with a $\phi \to K^+ K^-$ decay from the $K^+ K^- K_S^0$ sample.
The intermediate meson states are reconstructed from the following
decays: $\pi^0 \to \gamma \gamma$, $K_S^0 \to \pi^+ \pi^-$, 
$\omega \to \pi^+ \pi^- \pi^0$ and $f_0 (980) \to \pi^+ \pi^-$.
Charged tracks reconstructed with the CDC, except for tracks from 
$K_S^0 \to \pi^+ \pi^-$ decays, are required to originate from the 
interaction point (IP). We distinguish charged kaons from pions 
based on a kaon (pion) likelihood ${\cal L}_{K(\pi)}$ derived from 
the TOF, ACC and $dE/dx$ measurements in the CDC. Photons are identified
as isolated ECL clusters that are not matched to any charged track.
\par
\begin{figure}[htbp]
\begin{center}
\begin{tabular}{ccc}
\includegraphics[width=0.23\textwidth]{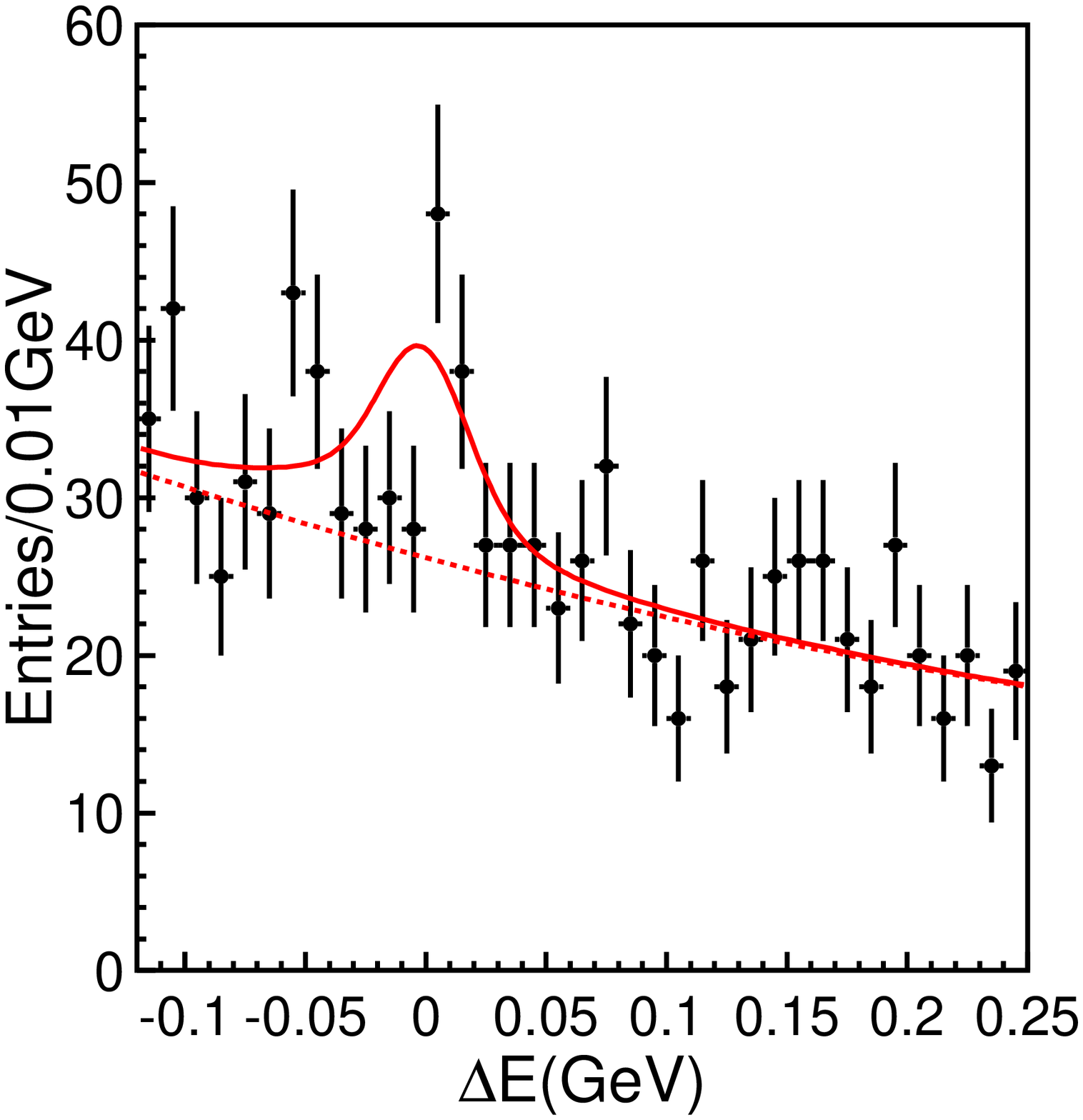} &
\includegraphics[width=0.23\textwidth]{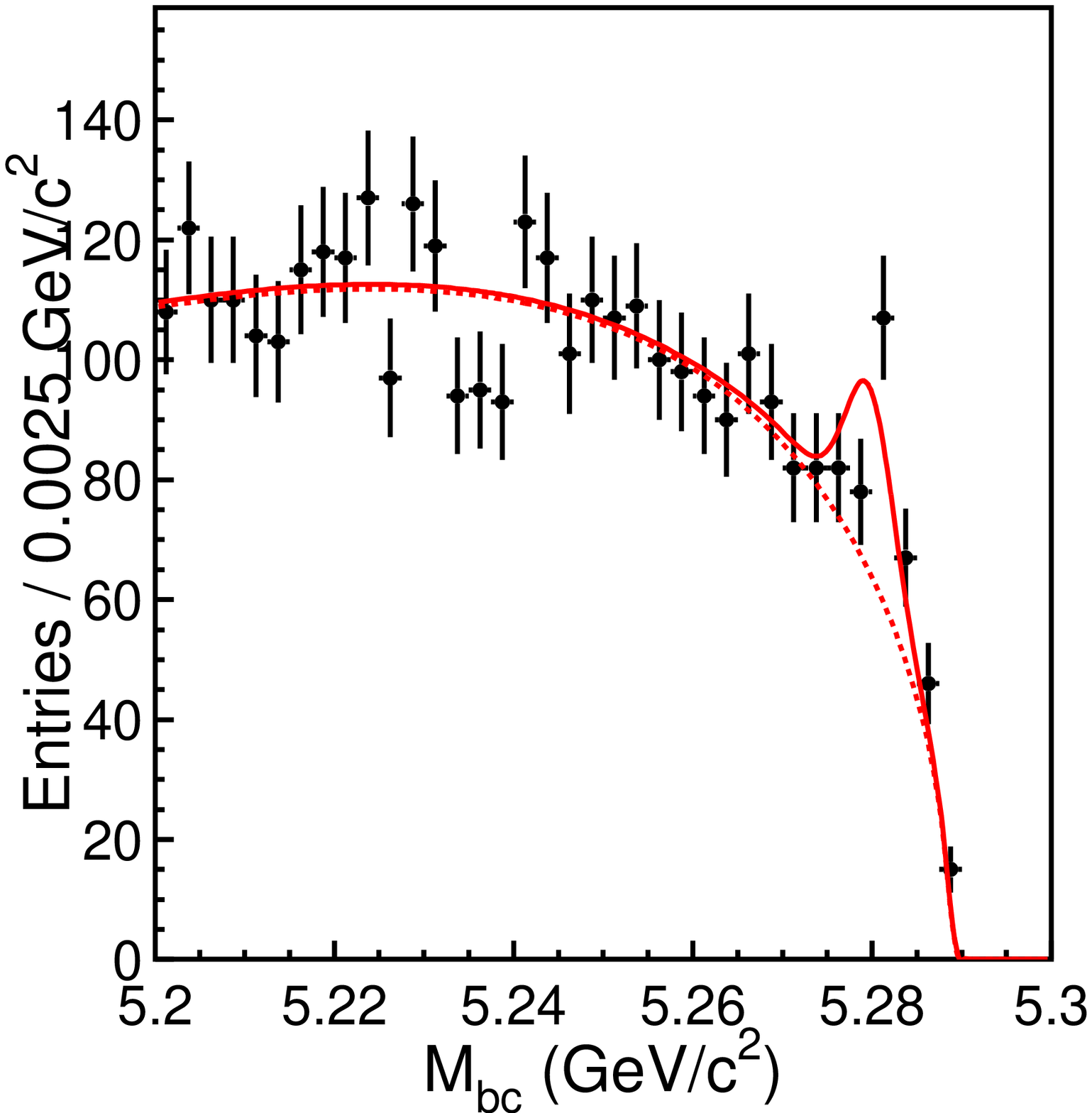} &
\includegraphics[width=0.23\textwidth]{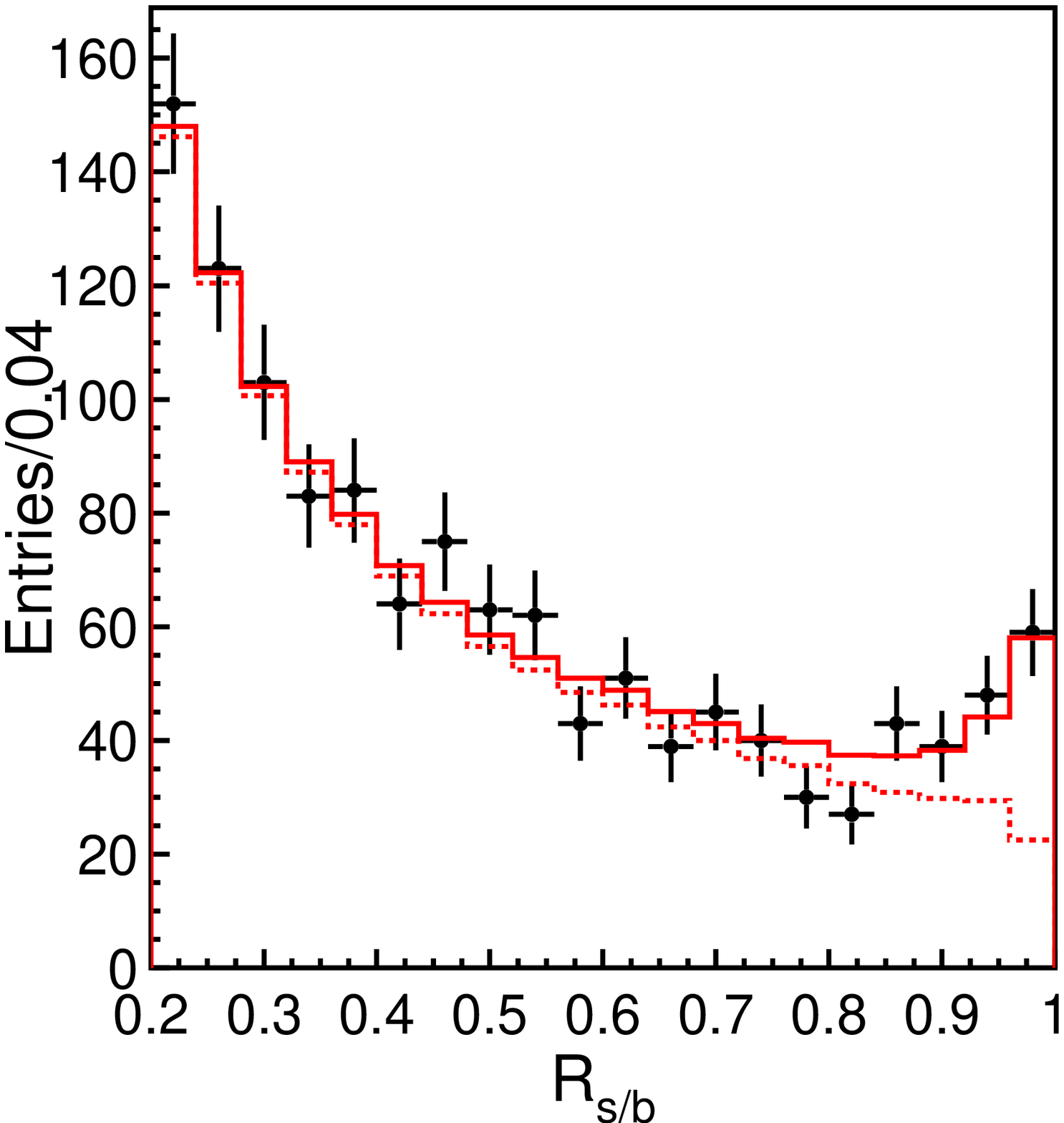} \\
\includegraphics[width=0.23\textwidth]{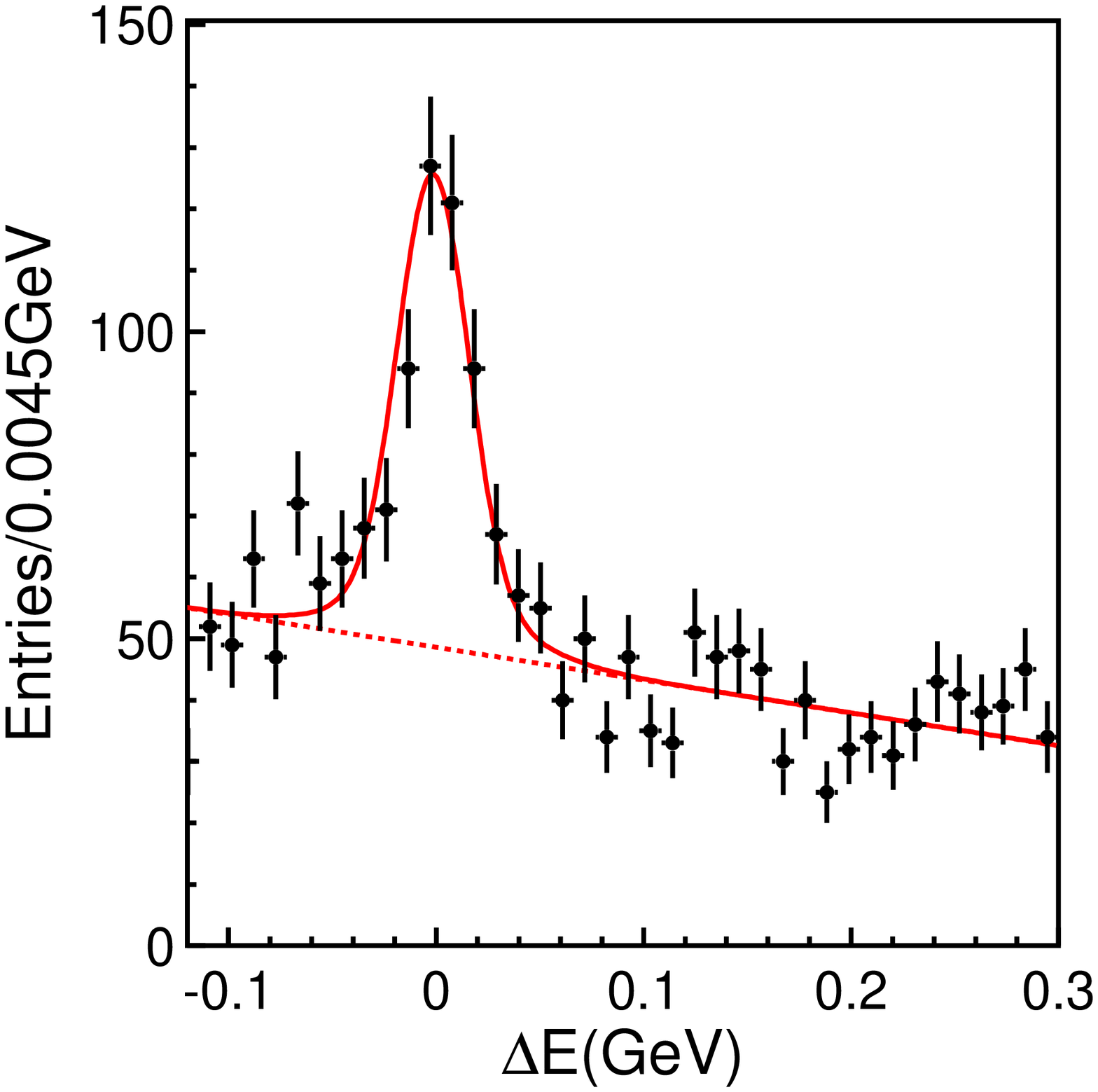} &
\includegraphics[width=0.23\textwidth]{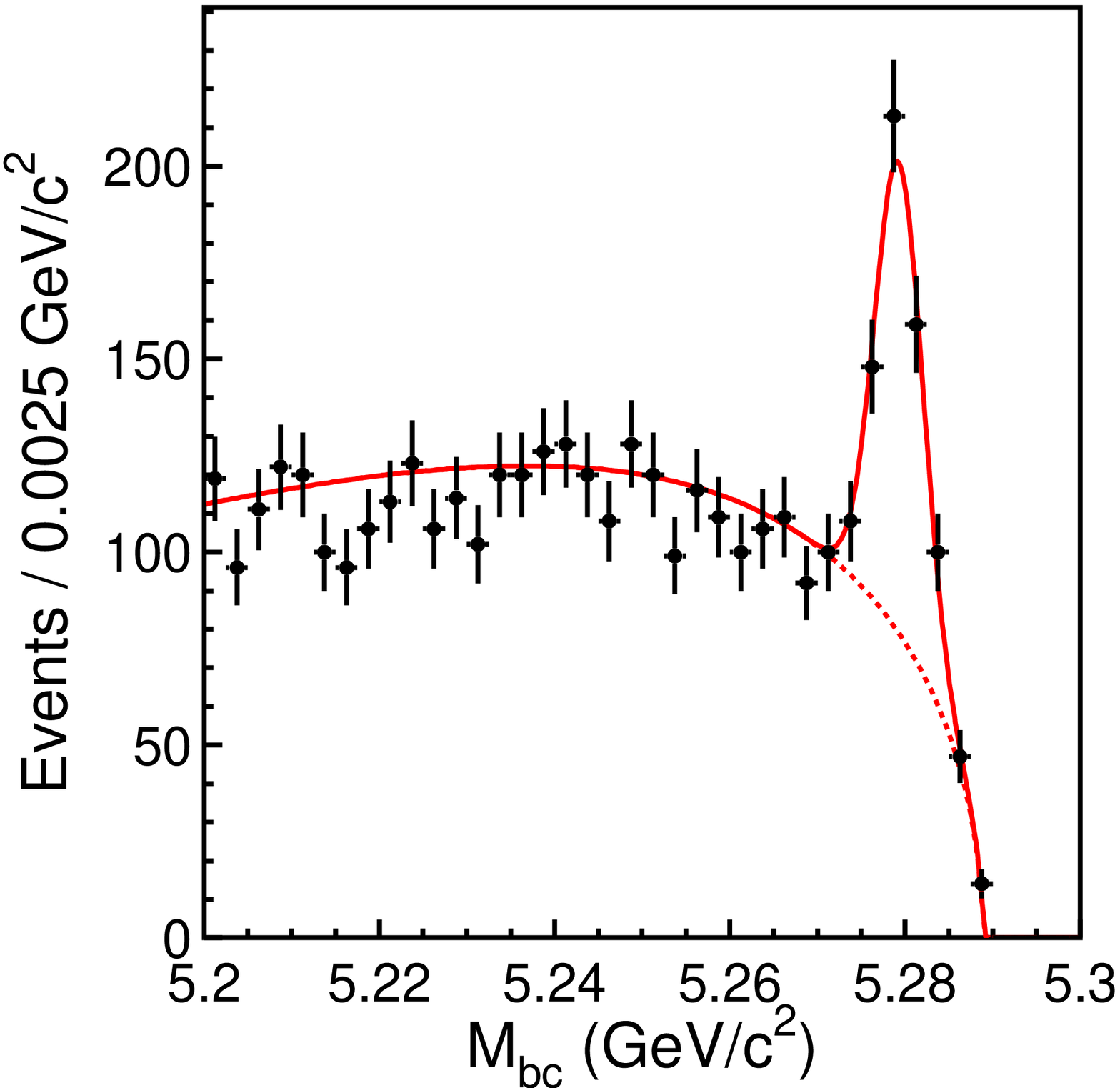} &
\includegraphics[width=0.23\textwidth]{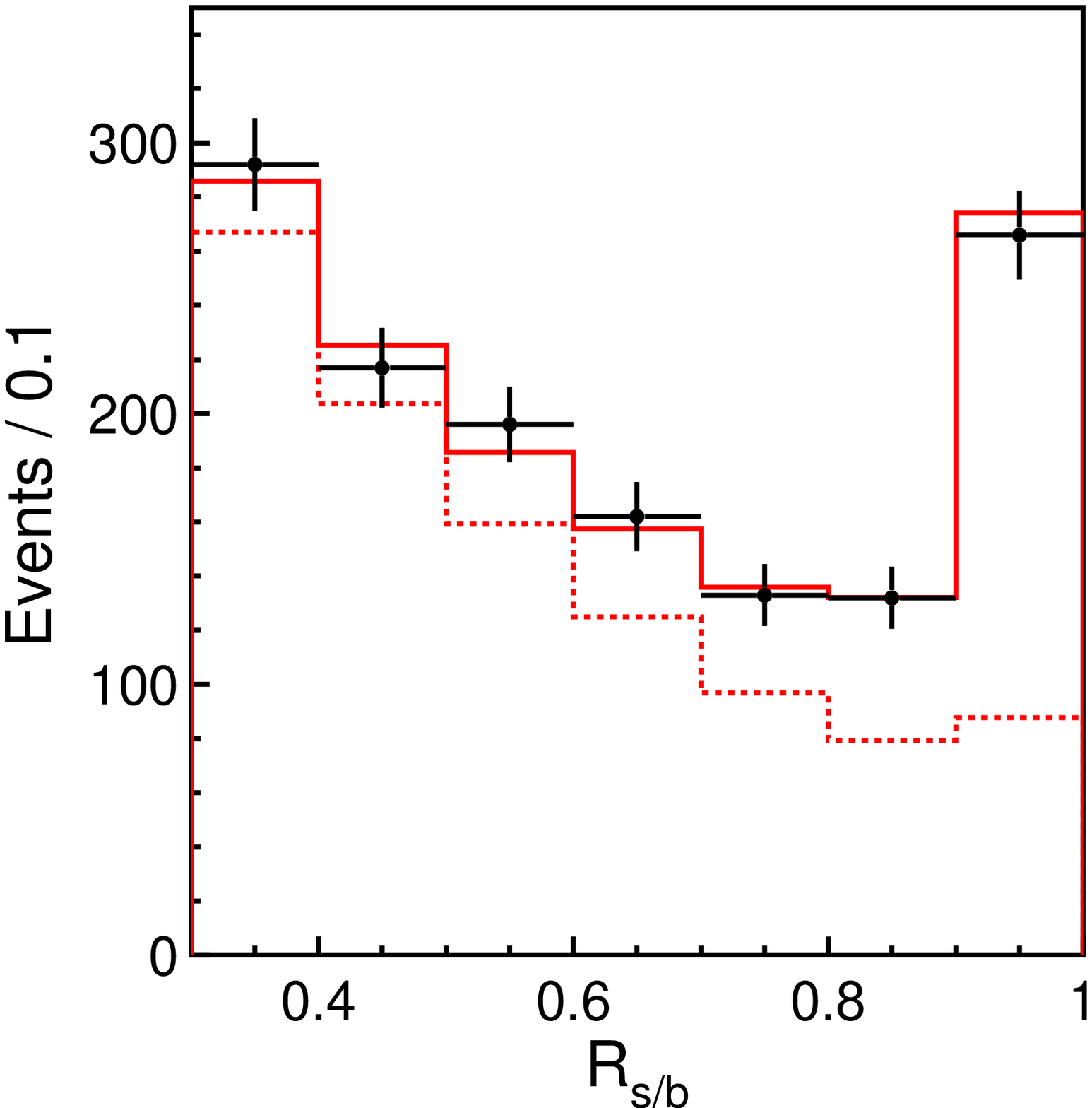} \\
\includegraphics[width=0.23\textwidth]{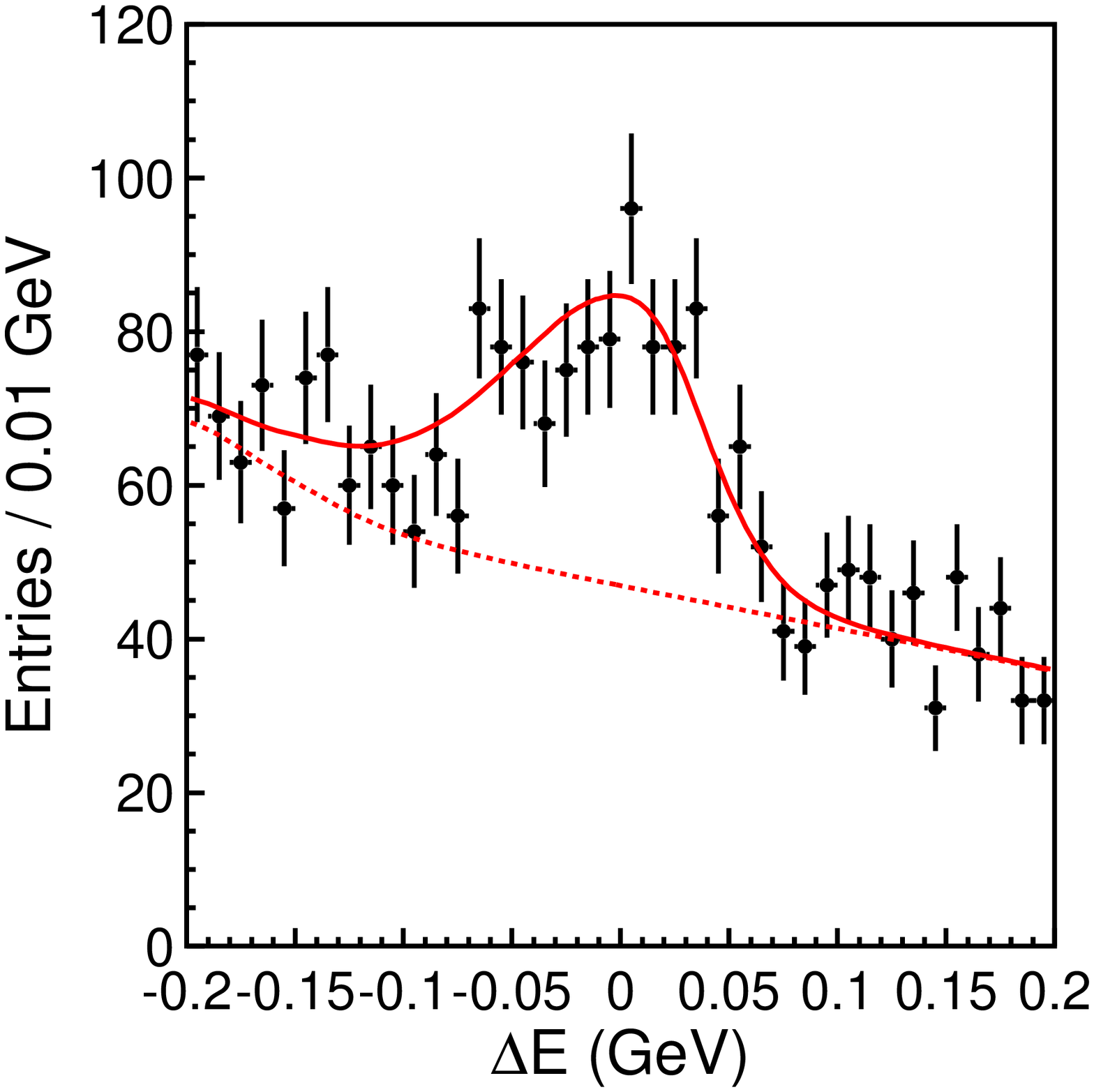} &
\includegraphics[width=0.23\textwidth]{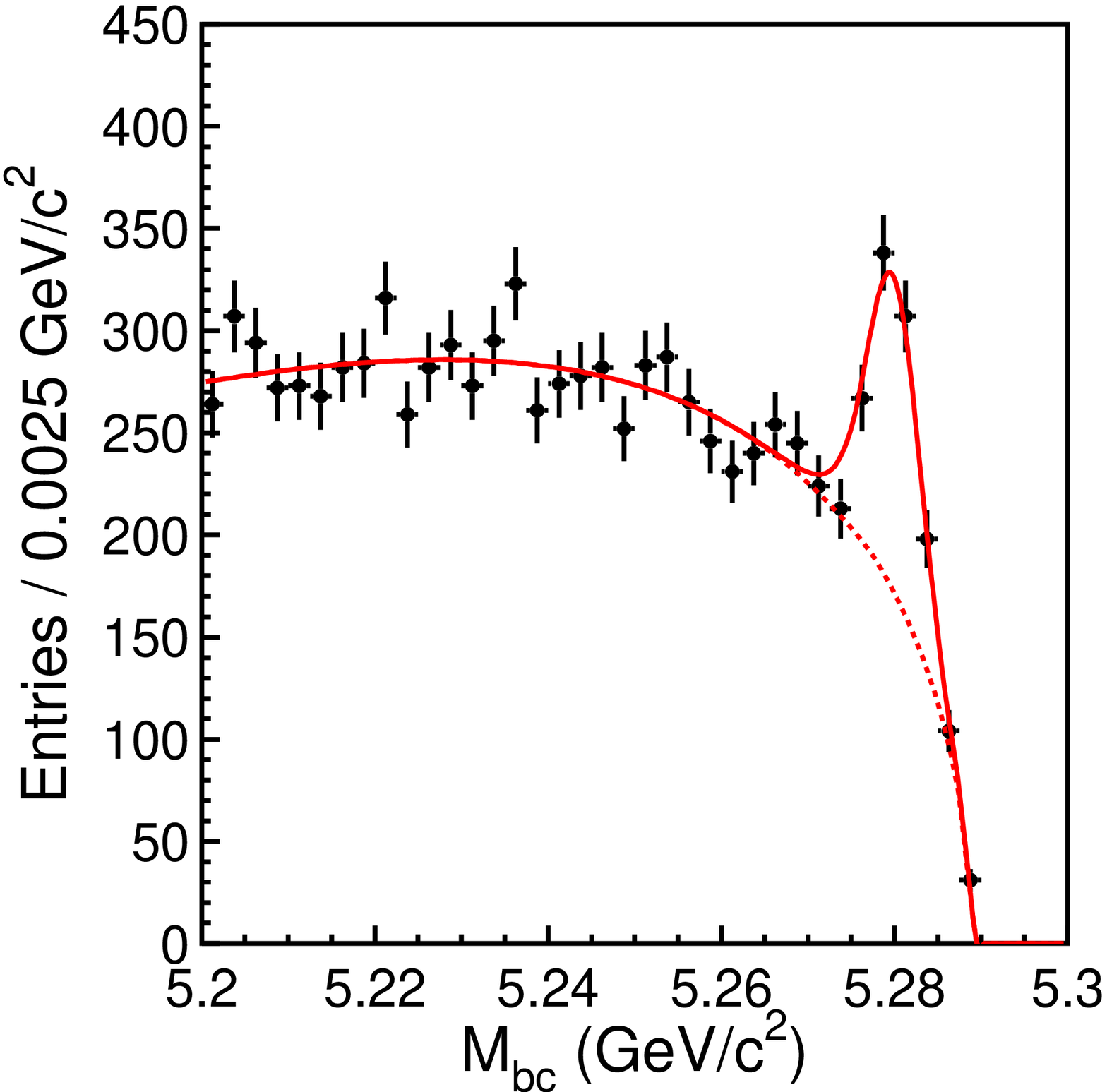} &
\includegraphics[width=0.23\textwidth]{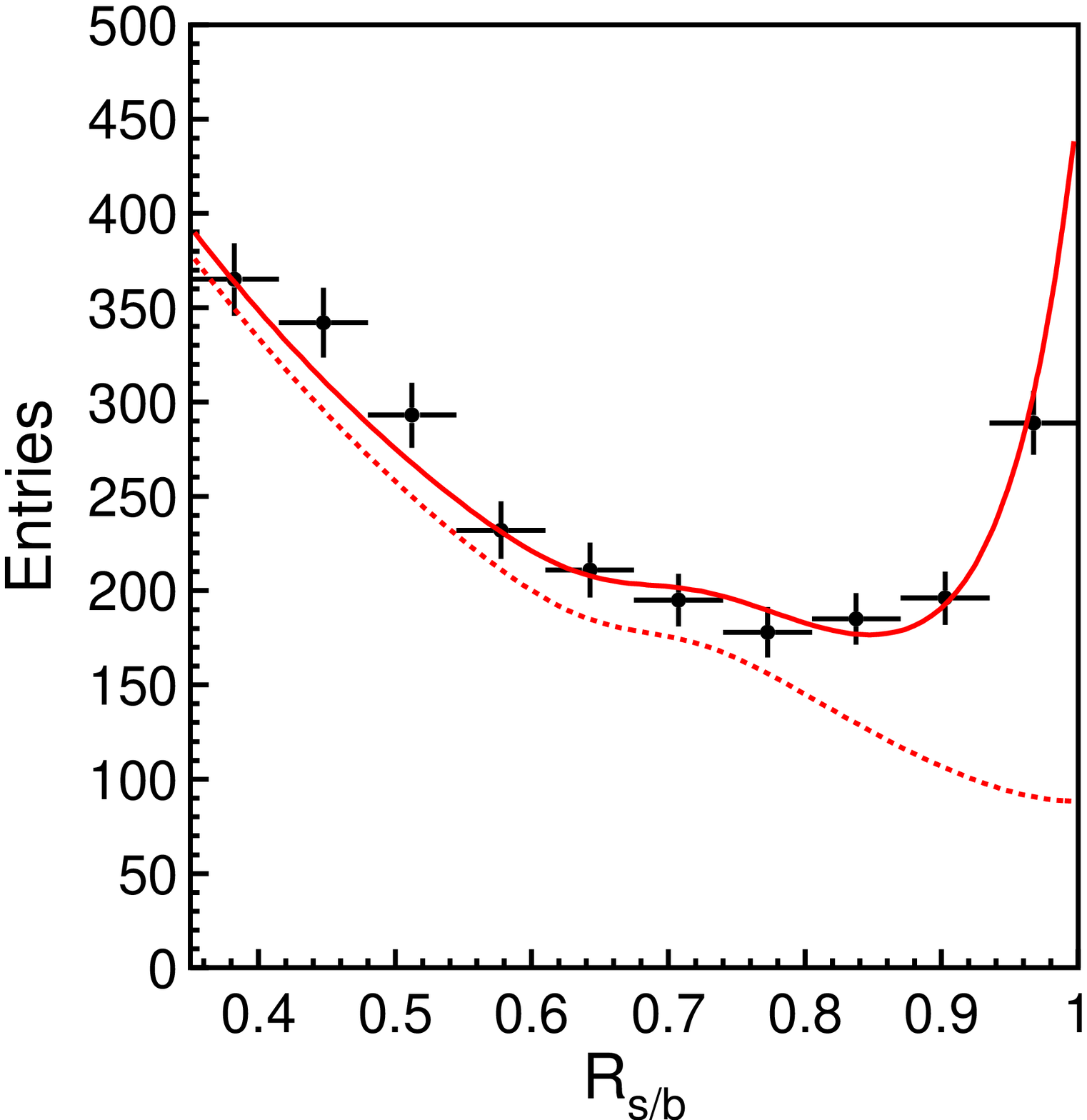} \\
\includegraphics[width=0.245\textwidth]{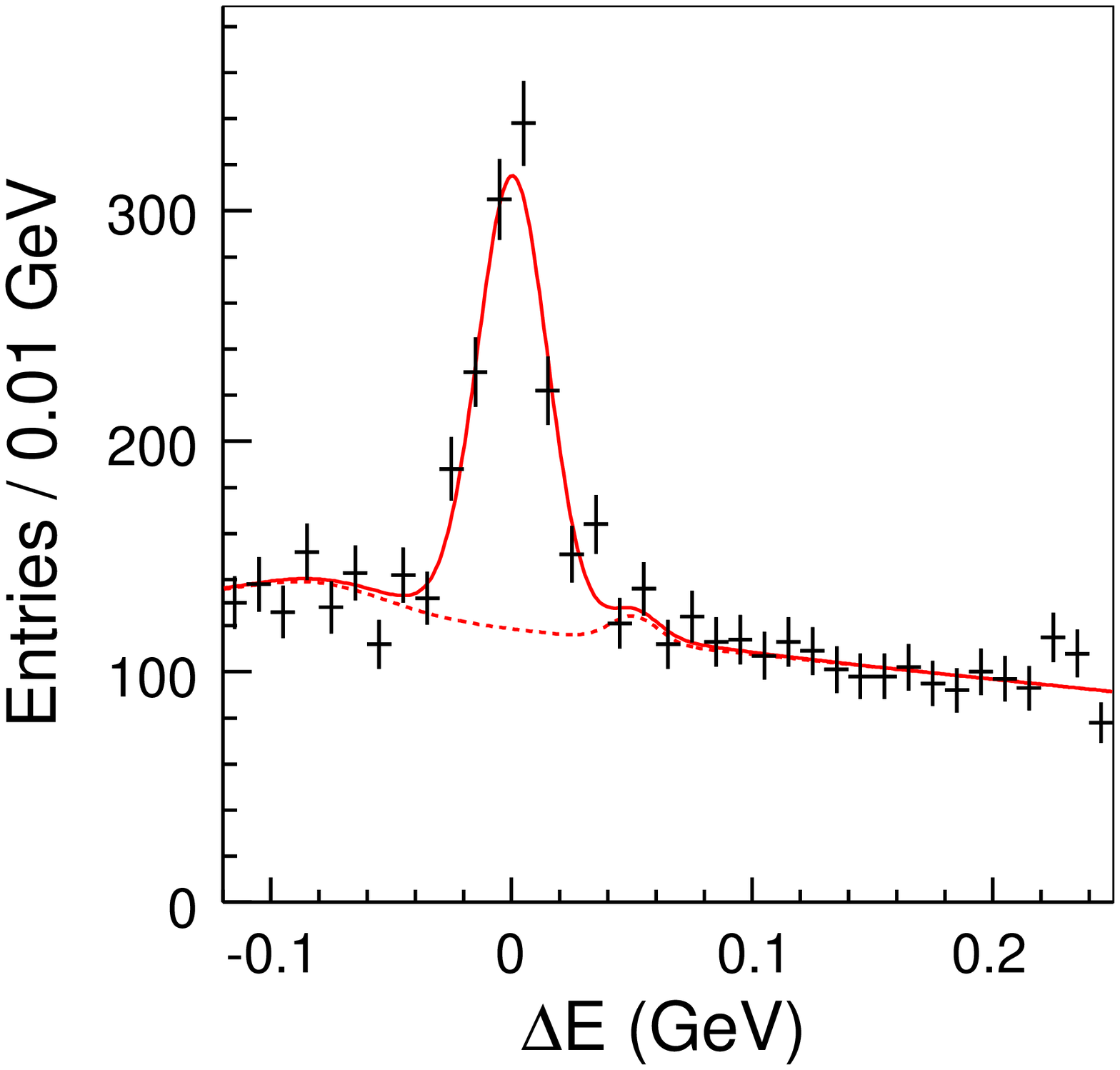} &
\includegraphics[width=0.245\textwidth]{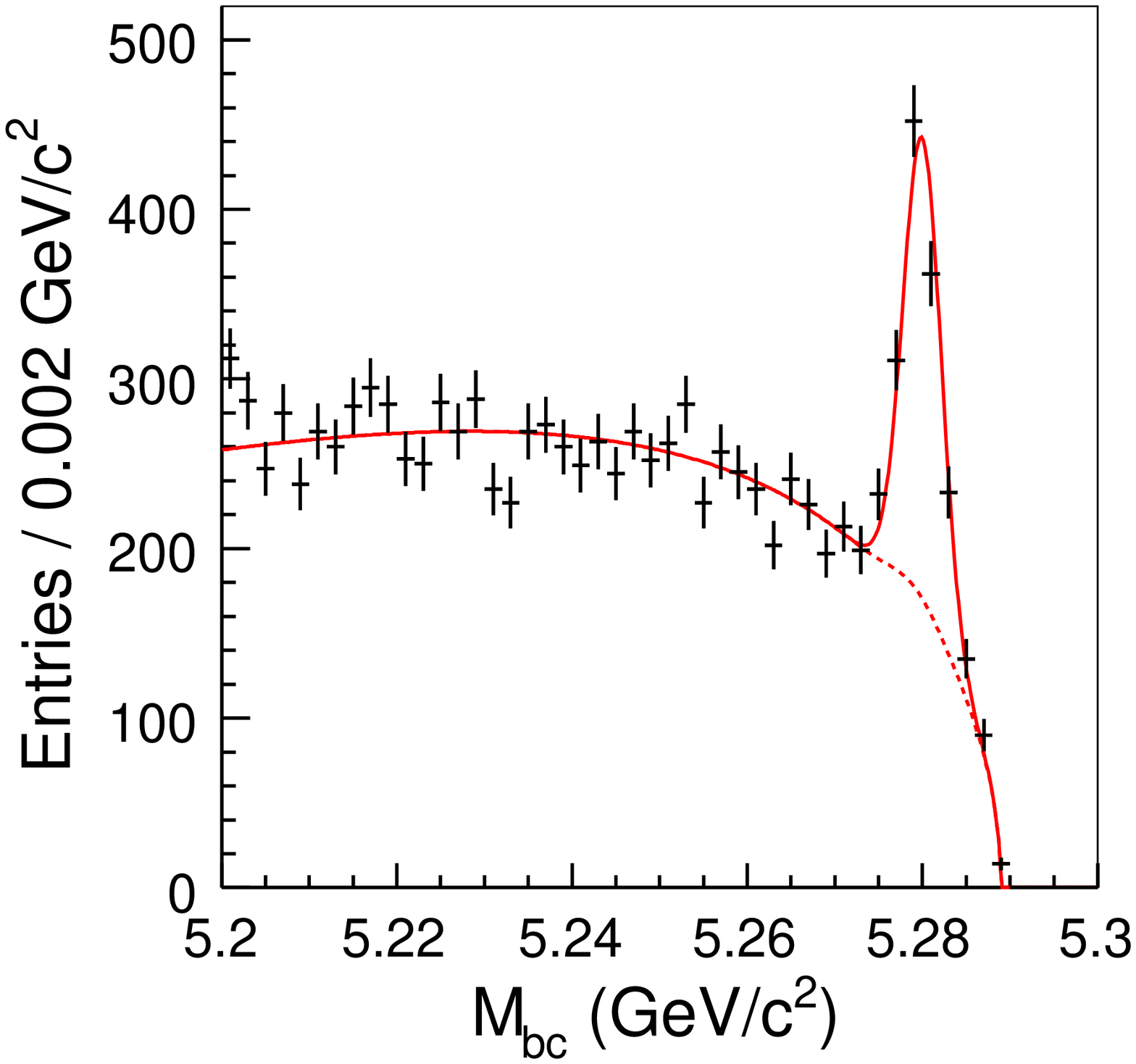} &
\includegraphics[width=0.245\textwidth]{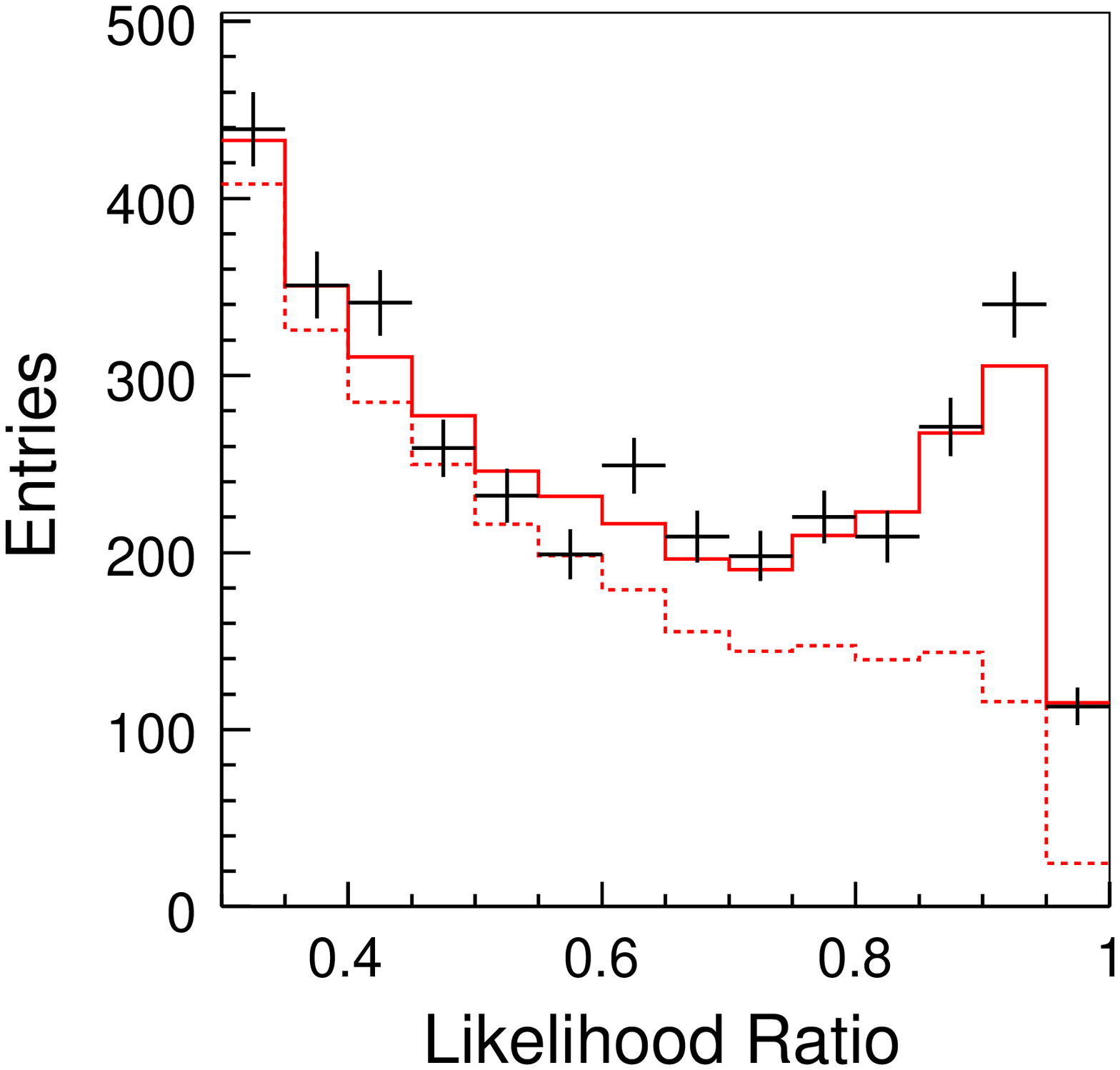} \\
\end{tabular}
\end{center}
\caption{$\Delta E$, $M_{\rm bc}$ and ${\cal R}_{\rm s/b}$ distributions for 
(a, b, c) $B^0 \to \omega K_S^0$, 
(d, e, f) $B^0 \to f_0 K_S^0$, (g, h, i) $B^0 \to K_S^0 \pi^0$ and 
(j, k, l) $B^0 \to K^+ K^- K_S^0$. The solid curves show the fits
to signal plus background distributions, and the dashed curves show the
background contributions.
To enhance the signal in the $\Delta E$ and $M_{\rm bc}$ projections, 
an additional cut on ${\cal R}_{\rm s/b}$ was applied ($> 0.5$).}
\rput[l]( -5.3, 18.4) {(a)}
\rput[l]( -1.1, 18.4) {(b)}
\rput[l]( +3.3, 18.4) {(c)}
\rput[l]( -5.3, 14.5) {(d)}
\rput[l]( -1.1, 14.5) {(e)}
\rput[l]( +3.3, 14.5) {(f)}
\rput[l]( -5.3, 10.5) {(g)}
\rput[l]( -1.1, 10.5) {(h)}
\rput[l]( +3.3, 10.5) {(i)}
\rput[l]( -5.3, 6.5) {(j)}
\rput[l]( -1.1, 6.5) {(k)}
\rput[l]( +3.3, 6.5) {(l)}
\label{fig_yield}
\end{figure}

 We identify $B$ meson decays using the energy difference 
$\Delta E \equiv E_B^{\rm cms} - E_{\rm beam}^{\rm cms}$ and the
beam-energy constrained mass 
$M_{\rm bc} \equiv \sqrt{(E_{\rm beam}^{\rm cms})^2 - (p_B^{\rm cms})^2}$,
where $E_{\rm beam}^{\rm cms}$ is the beam energy in the cms, and 
$E_B^{\rm cms}$ and $p_B^{\rm cms}$ are the cms energy and momentum 
of the reconstructed $B$ candidate, respectively. The signal candidates
are selected by requiring 5.27 GeV/$c^2 < M_{\rm bc} <$ 5.29 GeV/$c^2$
and a mode-dependent $\Delta E$ window. 
The dominant background for the $b \to s \overline{q} q$ signal comes from 
continuum events ($e^+ e^- \to u\overline{u}, d\overline{d}, s\overline{s}, c\overline{c}$).
We discriminate against this using event topology: continuum events
tend to be jet-like in the cms, while $e^+ e^- \to B\overline{B}$ events 
tend to be spherical. To quantify event topology, we calculate 
modified Fox-Wolfram moments
and combine them into a Fisher discriminant~\cite{SFW}. We calculate a 
probability density function (PDF) for this discriminant and multiply it by 
a PDF for $\cos \theta_B$, where $\theta_B$ is the polar angle in the cms
between the $B$ direction and the beam axis. The PDFs for signal and 
continuum are
obtained from Monte Carlo (MC) simulation and a data sideband, respectively. 
These PDFs are then used to calculate a signal [background] likelihood 
${\cal L}_{\rm sig[bkg]}$, and we impose loose mode-dependent requirements
on the likelihood ratio 
${\cal R}_{\rm s/b} \equiv {\cal L}_{\rm sig}/({\cal L}_{\rm sig}+
{\cal L}_{\rm bkg})$. 
Figures~\ref{fig_yield}.(a)-(l) show the reconstructed variables $M_{\rm bc}$, 
$\Delta E$ and ${\cal R}_{\rm s/b}$ after flavor tagging and vertex 
reconstruction 
(before vertex reconstruction for the decay $B^0 \to K_S^0 \pi^0$);
the corresponding signal yields are summarized in Table~\ref{tab_yield}.
\begin{table}[htbp]
\caption{Estimated signal yields $N_{\rm sig}$ in the signal
region for each mode.}
\begin{center}
\begin{tabular}{lcc}
\hline
\hline
Mode  & $\xi_f$ & $N_{\rm sig}$\\
\hline
$\omega K_S^0$ & $-1$ & $118 \pm 18$\\
$f_0 K_S^0$ & $+1$ & $377 \pm 25$ \\
$K_S^0 \pi^0$ & $-1$ & $515 \pm 32$ \\
$K^+ K^- K_S^0$ & $\;\;\;+0.86 \pm 0.18 \pm 0.09\;\;\;$ & $840 \pm 34$ \\
\hline
\hline
\end{tabular}
\end{center}
\label{tab_yield}
\end{table}
%
%
\par We determine ${\cal S}_f$ and ${\cal A}_f$ for each mode by performing an unbinned
maximum-likelihood fit to the observed $\Delta t$ distribution.
The decay rate is given by 
\begin{eqnarray}
\label{eq_decay}
{\cal P}(\Delta{t}) = \frac{ e^{-|\Delta{t}|/{\tau_{B^0}}} }{4\tau_{B^0}}
\biggl\{1 + q\cdot 
\Bigl[ {\cal S}_f \sin(\Delta m_d \Delta{t}) 
   + {\cal A}_f \cos(\Delta m_d \Delta{t})
\Bigr]
\biggr\}
\end{eqnarray}
where $\tau_{B^0}$ is the $B^0$ lifetime and the $b$-flavor charge $q = +1 (-1)$
when the tagging $B$ meson is a $B^0$ ($\overline{B}{}^0$). 
The $b$-flavor of the accompanying $B$ meson is identified by a tagging 
algorithm~\cite{tag} that categorizes charged leptons, kaons and $\Lambda$'s found
in the event. The algorithm returns two parameters: the $b$-flavor charge $q$
and $r$, which indicates the tag quality as determined from MC
simulation and varies from $r=0$ for no flavor discrimination to $r = 1$ for 
unambiguous flavor assignment. 
If $r \leq 0.1$, we set the wrong tag fraction to 0.5, and therefore the 
accompanying $B$ meson provides no tagging information in this case.
Events with $r > 0.1$ are sorted into six intervals.
\par To the PDF expected for the signal distribution (Eq.~\ref{eq_decay}), 
the effect of incorrect flavor assignment is incorporated and 
then convolved with a resolution function $R_{\rm sig}(\Delta t)$ to take 
into account the finite vertex resolution.
The wrong tag fractions for the six $r$ intervals,
$w_l$ ($l = 1, 6$), and differences between $B^0$ and $\overline{B}{}^0$
decays, $\Delta w_l$, as well as the resolution parameters
are determined using a high-statistics control
sample of semileptonic and hadronic $b \to c$ decays.
\par
\begin{figure}[htbp]
\includegraphics[width=0.75\textwidth]{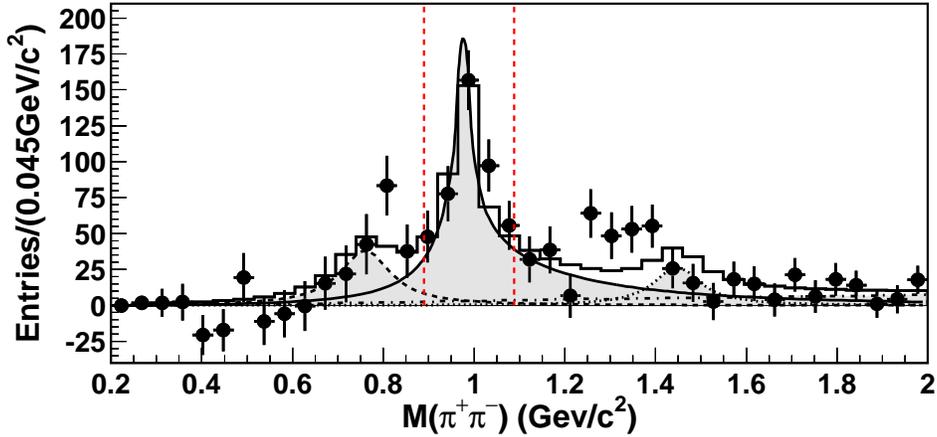}
\caption{$\pi^+ \pi^-$ mass distribution for the $f_0 K_S$ events in 
the $\Delta E$-$M_{\rm bc}$ signal box (shown here after background 
subtraction). The histogram is the result of the fit whereas the different 
contributions are 
shown (continuous line for $f_0 (980)$, dashed for $\rho^0$ and dotted for 
$f_X$).}
\label{fig_f0ks}
\end{figure}

 We determine the following likelihood for each event:
\begin{eqnarray}
P_i
&=& (1-f_{\rm ol}) \int 
\biggl[
f_{\rm sig} {\cal P}_{\rm sig}(\Delta t') R_{\rm sig}(\Delta t_i-\Delta t') \nonumber \\
&+&(1-f_{\rm sig}){\cal P}_{\rm bkg}(\Delta t')R_{\rm bkg}(\Delta t_i-\Delta t')\biggr]
d(\Delta t')  \nonumber \\
&+&f_{\rm ol} P_{\rm ol}(\Delta t_i).
\label{eq:likelihood}
\end{eqnarray}
The signal probability $f_{\rm sig}$ depends on the $r$ region and is 
calculated on an event-by-event basis as a function of 
$M_{\rm bc}$, $\Delta E$ and ${\cal R}_{\rm s/b}$ (and 
$M(\pi^+ \pi^- \pi^0)$ for $B^0 \to \omega K_S^0$). 
For $B^0 \to f_0 (980) K_S^0$, this fit yields the number of 
$B^0 \to \pi^+ \pi^- K_S^0$ candidates that have $\pi^+ \pi^-$ invariant 
mass within the $f_0 (980)$ resonance region, which includes other 
contributions (e.g. $B^0 \to \rho^0 K_S^0$, 
$K^* \pi^\pm$ and non-resonant three-body decays) which peak 
like the signal in $\Delta E$ and $M_{\rm bc}$ distributions. 
To estimate these peaking backgrounds, we perform a fit to the 
$\pi^+ \pi^-$ invariant 
mass distribution for the events inside the $\Delta E$-$M_{\rm bc}$ 
signal region (Fig~\ref{fig_f0ks}).

The PDF for background events, ${\cal P}_{\rm bkg}(\Delta t)$, is modeled as 
a sum of exponential and prompt components and is convolved with a sum of
two Gaussians, $R_{\rm bkg}$. Parameters in ${\cal P}_{\rm bkg}(\Delta t)$
and $R_{\rm bkg}$ for background are determined from a fit 
to the $\Delta t$ distribution for events in the $\Delta E$-$M_{\rm bc}$ 
data sideband. $P_{\rm ol}(\Delta t)$ is a broad Gaussian function that 
represents an outlier component with a small fraction $f_{\rm ol}$. 
The only free parameters in the final fits are ${\cal S}_f$ and 
${\cal A}_f$, which are determined by maximizing the likelihood 
function $L = \prod_i P_i(\Delta t_i; {\cal S}_f, {\cal A}_f)$ where 
the product is over all events.
%
\par
\begin{figure}[htbp]
\begin{center}
\begin{tabular}{cc}
\includegraphics[width=0.4\textwidth]{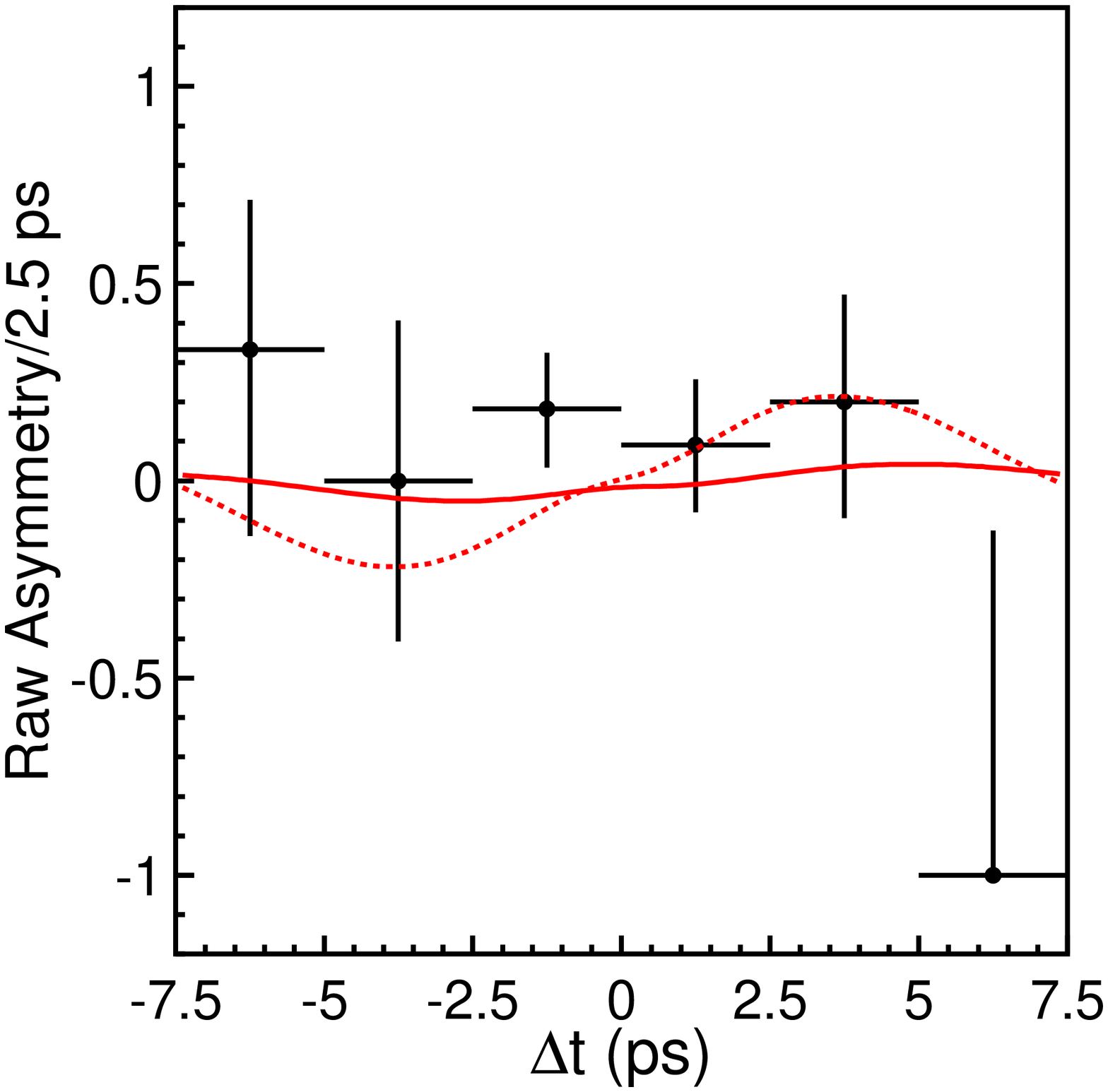} & 
\includegraphics[width=0.4\textwidth]{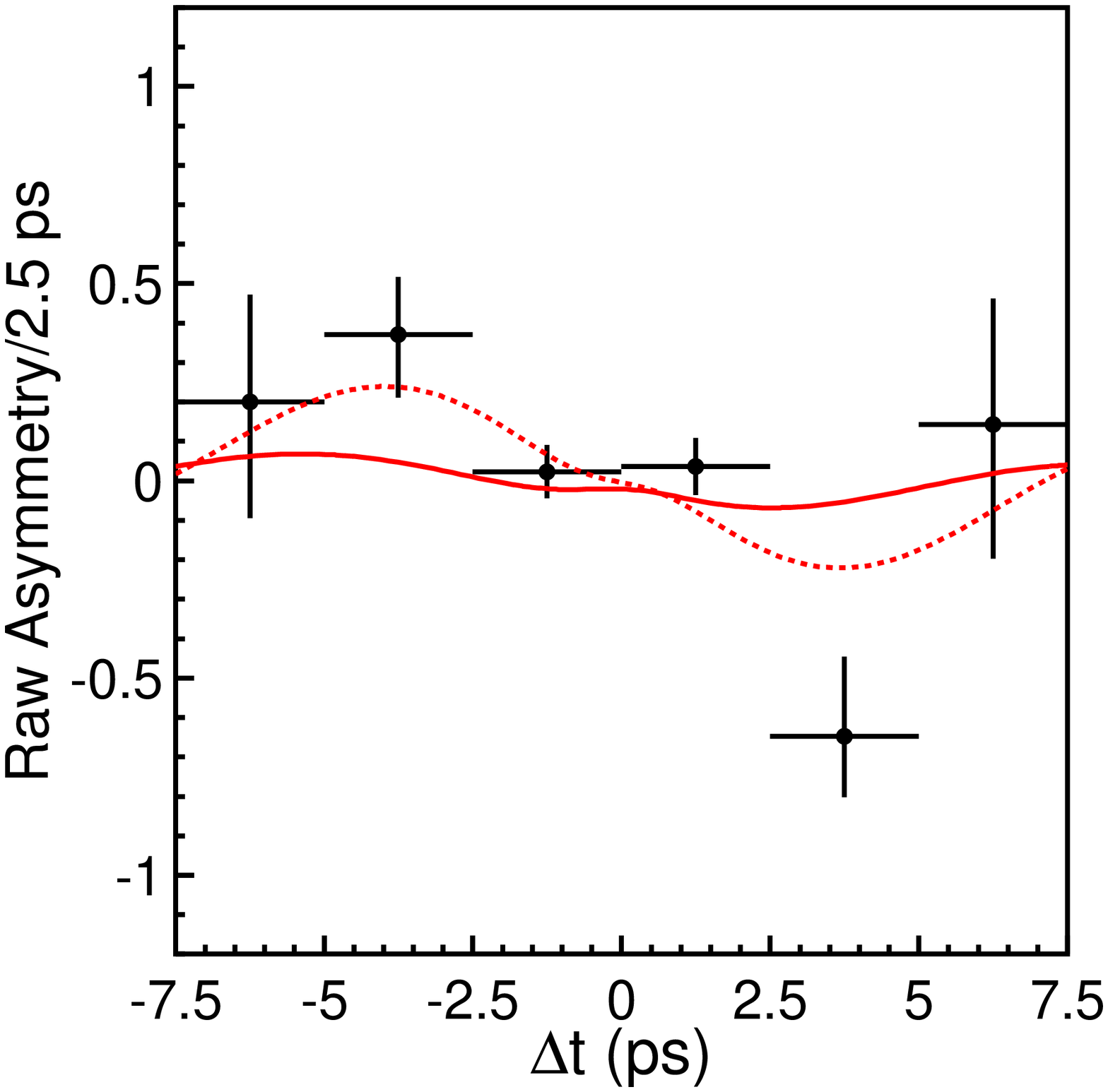} \\
\includegraphics[width=0.4\textwidth]{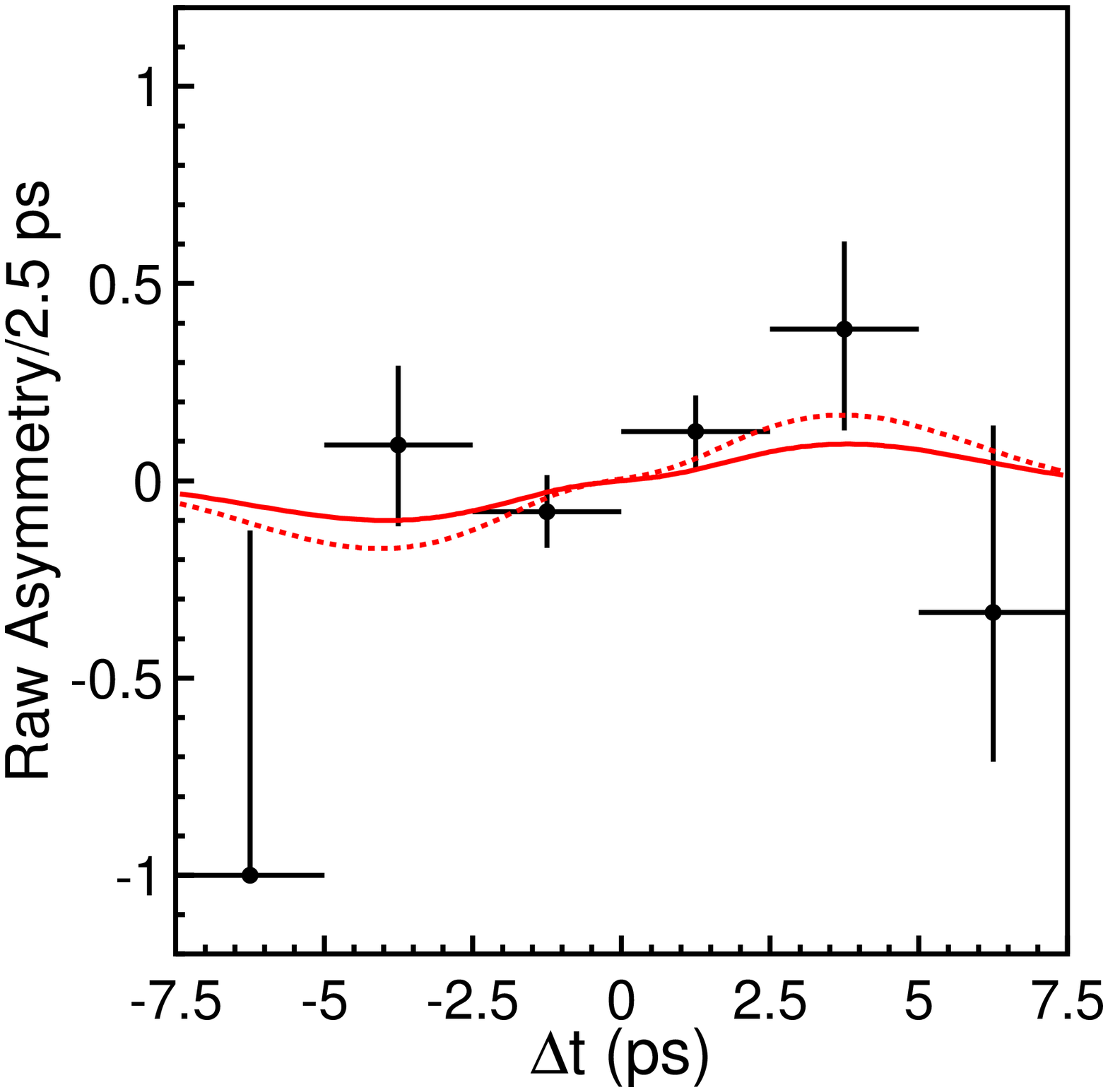} & 
\includegraphics[width=0.43\textwidth]{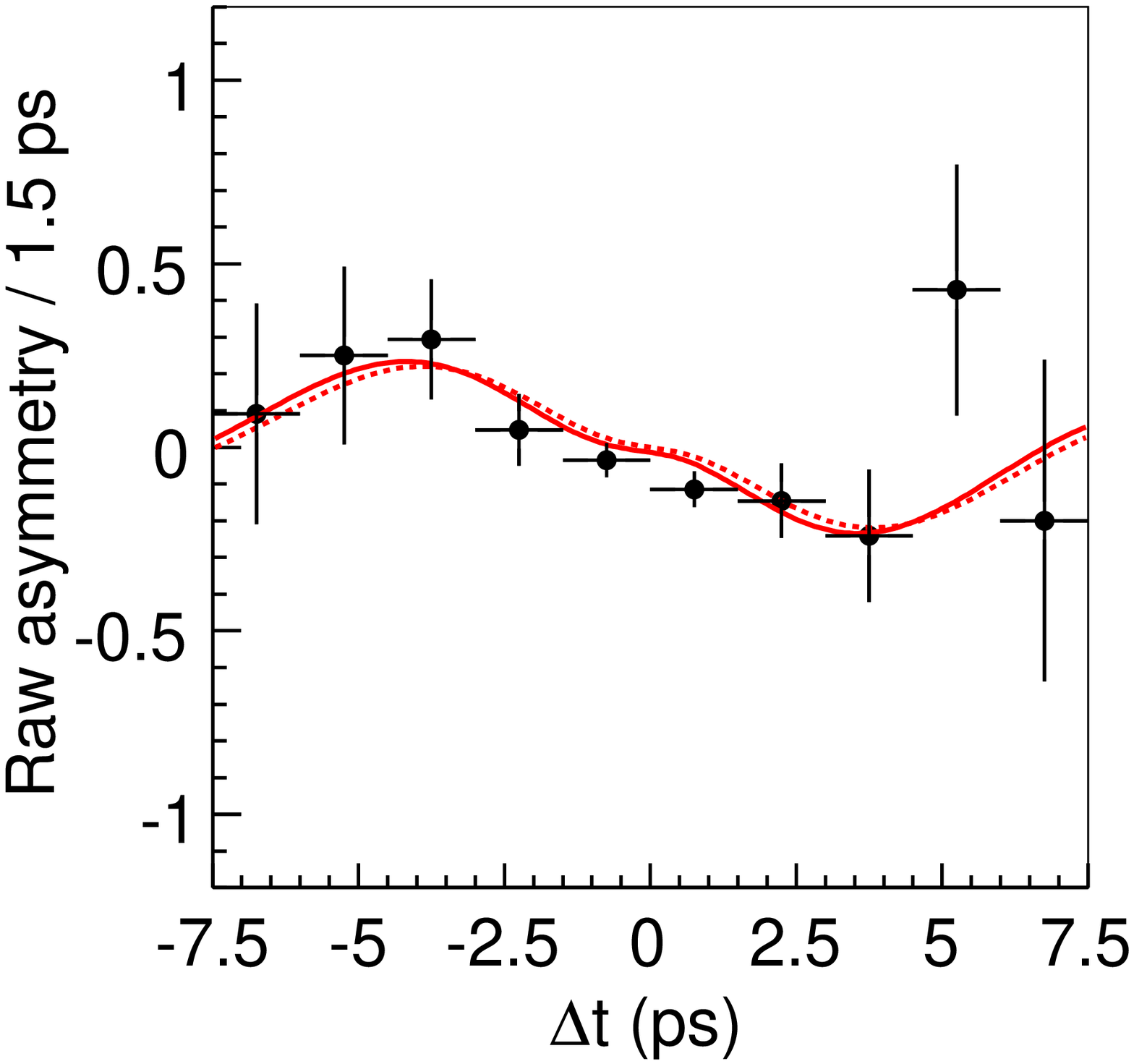} 
\end{tabular}
\end{center}
\caption{Asymmetries
of good-tagged events ($r > 0.5$) for (a) $B^0 \to \omega K_S^0$, 
(b) $B^0 \to f_0(980) K_S^0$, (c) $B^0 \to K_S^0 \pi^0$ and 
(d) $B^0 \to K^+ K^- K_S^0$. The solid curves show the results
of the unbinned maximum-likelihood fits. The dashed curves show the
SM expectation with the measurement of $CP$-violation parameters 
for the $B^0 \to J/\psi K^0$ mode ($\sin 2 \phi_1 = + 0.642$ and 
${\cal A}_f = 0$)}
\rput[l]( -5.5, 15.65) {(a)}
\rput[l]( +1.5, 15.65) {(b)}
\rput[l]( -5.5, 8.8) {(c)}
\rput[l]( +1.5, 8.8) {(d)}
\label{raw_asym}
\end{figure}

 Table~\ref{sin2phi1} summarizes the fit results of 
$\sin 2 \phi_1^{\rm eff}$ and ${\cal A}_f$. 
For the $B^0 \to K^+ K^- K_S^0$ decay, the SM prediction is given 
by ${\cal S}_f = -(2f_+ -1) \sin 2 \phi_1^{\rm eff}$. 
The effective $\sin 2 \phi_1$ value for this mode is found to be 
$+0.68 \pm 0.15 \pm 0.03^{+0.21}_{-0.13}$. The third error is an
additional systematic error arising from the uncertainty of the
$\xi_f = +1$ fraction.
We define the raw asymmetry in each $\Delta t$ bin by 
$(N_{q=+1}-N_{q=-1})/(N_{q=+1}+N_{q=-1})$, where $N_{q=+1(-1)}$
is the number of observed candidates with $q=+1(-1)$.
Figure~\ref{raw_asym} shows this asymmetry
for good tag quality ($r > 0.5$) events in each mode.
\begin{table}[htbp]
\caption{Results of the fits to the $\Delta t$ distributions. The first error
is statistical and the second error is systematic. The third error for
$\sin 2 \phi_1^{\rm eff}$ of $K^+ K^- K_S^0$ is an
additional systematic error arising from the uncertainty of the
$\xi_f = +1$ fraction.}
\begin{center}
\begin{tabular}{lcc}
\hline
\hline
Mode  & $\sin 2 \phi_1^{\rm eff}$ & ${\cal A}_f$\\
\hline
$\omega K_S^0$  & $+0.11 \pm 0.46 \pm 0.07$ & $-0.09 \pm 0.29 \pm 0.06$ \\
$f_0 K_S^0$     & $+0.18 \pm 0.23 \pm 0.11$ & $-0.15 \pm 0.15 \pm 0.07$ \\
$K_S^0 \pi^0$   & $+0.33 \pm 0.35 \pm 0.08$ & $-0.05 \pm 0.14 \pm 0.05$ \\
$K^+ K^- K_S^0$ & $\;\;\;+0.68 \pm 0.15 \pm 0.03^{+0.21}_{-0.13}\;\;\;$ & 
$-0.09 \pm 0.10 \pm 0.05$ \\
\hline
\hline
\end{tabular}
\end{center}
\label{sin2phi1}
\end{table}
The dominant sources of systematic error for ${\cal S}_f$ in 
$b \to s \overline{q} q$ modes are
the uncertainties in the vertex reconstruction (0.01),
in the background fraction (from 0.01 for $K_S \pi^0$ to 0.04 in $\omega K_S^0$)
and in the background $\Delta t$ distribution 
(0.04 in $K_S \pi^0$ and 0.01 or less in others), 
and in the resolution function (0.05 for $\omega K_S$ and $K_S \pi^0$).
The dominant sources for ${\cal A}_f$ are the effects of tag-side 
interference~\cite{long} (0.04),
the uncertainties in the vertex reconstruction (0.02),
in the background fraction (0.03 for $f_0 K_S^0$ and $\omega K_S^0$ and $< 0.02$
for others).
For the $f_0 K_S^0$ mode, additional systematics were included: uncertainties 
from the $M(\pi \pi)$ fit (0.06 for ${\cal S}_f$) and from the assumption on 
the $CP$ content of the peaking background (0.08 for ${\cal S}_f$ and 0.04 
for ${\cal A}_f$). 
For the $K_S^0 \pi^0$ mode, the uncertainty on the rare $B$ component is 
taken into account (0.04 for ${\cal S}_f$ and 0.02 for ${\cal A}_f$). 
Other contributions come from uncertainties in wrong tag fractions, 
lifetime and mixing.
A possible fit bias is examined by fitting a large number of MC events.
We add each contribution in quadrature to obtain the total systematic 
uncertainty.
%
\par In summary, we have performed improved measurements of $CP$-violation
parameters $\sin 2 \phi_1^{eff}$ and ${\cal A}_f$ for $B^0 \to \omega K_S^0,
f_0(980) K_S^0, K_S^0 \pi^0$ and $K^+ K^- K_S^0$
using 535 $\times 10^6$ $B\overline{B}$ events.\\
Comparing the results for each individual $b \to s $ mode with those from the
$B^0 \to J/\psi K^0$ decay, we have not observed a significant deviation with 
the present statistics. 

\section*{Acknowledgments}
We thank the KEKB group for the excellent operation of the
accelerator, the KEK cryogenics group for the efficient
operation of the solenoid, and the KEK computer group and
the National Institute of Informatics for valuable computing
and Super-SINET network support. We acknowledge support from
the Ministry of Education, Culture, Sports, Science, and
Technology of Japan and the Japan Society for the Promotion
of Science; the Australian Research Council and the
Australian Department of Education, Science and Training;
the National Science Foundation of China and the Knowledge
Innovation Program of the Chinese Academy of Sciencies under
contract No.~10575109 and IHEP-U-503; the Department of
Science and Technology of India; 
the BK21 program of the Ministry of Education of Korea, 
the CHEP SRC program and Basic Research program 
(grant No.~R01-2005-000-10089-0) of the Korea Science and
Engineering Foundation, and the Pure Basic Research Group 
program of the Korea Research Foundation; 
the Polish State Committee for Scientific Research; 
the Ministry of Science and Technology of the Russian
Federation; the Slovenian Research Agency;  the Swiss
National Science Foundation; the National Science Council
and the Ministry of Education of Taiwan; and the U.S.\
Department of Energy.

\end{document}